\documentstyle[12pt]{article}
\def\beq{\begin{equation}}
\def\eeq{\end{equation}}
\def\bea{\begin{eqnarray}}
\def\eea{\end{eqnarray}}
\def\bq{\begin{quote}}
\def\eq{\end{quote}}

\def\nnb{\nonumber}
\def\ga{\left(}
\def\dr{\right)}
\def\aga{\left\{}
\def\adr{\right\}}

\def\nnb{\nonumber}
\def\la{\langle}
\def\ra{\rangle}
\def\nin{\noindent}

\def\mr{\overline{m}}
\def\alf{\alpha_s}
\def\al{\Lambda}

\begin{document}
\pagestyle{empty}
\begin{flushright}
{CERN-TH.7405/94}\\
\end{flushright}
 \vspace*{5mm}
\begin{center}
\section*{\bf
A fresh look into the heavy quark-mass values }
{\bf S. Narison} \\
 \vspace{0.3cm}
Theoretical Physics Division, CERN\\
CH - 1211 Geneva 23, Switzerland\\
and\\
Laboratoire de Physique Math\'ematique\\
Universit\'e de Montpellier II\\
Place Eug\`ene Bataillon\\
34095 - Montpellier Cedex 05, France\\
\vspace*{1cm}
{\bf Abstract} \\ \end{center}
\vspace*{2mm}
\noindent
Using the recent {\it world average}
 $\alpha_s(M^2_Z)= 0.118 \pm 0.006$, we give the
 {\it first direct extraction}
from the $\Psi $ and $ \Upsilon$
data of the values of the
{\it running heavy quark masses} within QCD spectral
sum rules to two-loops in the $\overline {MS}$-scheme:
$\mr_b(M^{PT2}_b)$ = $(4.23~^{+0.03}_{-0.04} \pm 0.02)$ GeV
and $\mr_c(M^{PT2}_c)$ = $(1.23~^{+ 0.02}_{-0.04} \pm 0.03)$ GeV,
(the errors are respectively due to $\alf$ and to the
gluon condensate), and the corresponding value of
the {\it short-distance perturbative pole masses to two-loops}:
$M^{PT2}_b=(4.62 \pm 0.02)$ GeV,	 $M^{PT2}_c=(1.41\pm 0.03)$ GeV,
which we compare with the
 updated values of the {\it non-relativistic
pole masses} re-extracted {\it directly}
from the two-loop non-relativistic sum rules:
$M^{NR}_b= (4.69~^{-0.01}_{+0.02}\pm 0.02)$ GeV and
$M^{NR}_c=(1.44\pm 0.02\pm 0.03)$ GeV. It is also informative
to compare the {\it three-loop} values of the
short-distance pole masses:
$M^{PT3}_b=(4.87\pm 0.05
 \pm 0.02)$ GeV and $M^{PT3}_c=(1.62\pm 0.07 \pm 0.03)$ GeV, with the
{\it dressed mass} $M^{nr}_b =(4.94 \pm 0.10 \pm 0.03)$ GeV,
entering into
the {\it non-relativistic Balmer formula} including higher order $\alf$
corrections. The $small$ mass-differences $M^{NR}_b-M^{PT2}_b
\simeq M^{nr}_b-M^{PT3}_b \simeq 70$ MeV and $M_c^{NR}-M_c^{PT2} \simeq
(30 \pm 20)$ MeV {\it can measure the size}
of the non-perturbative effect induced by {\it renormalon}
type-singularities. An analogous analysis
is pursued for the heavy-light mesons,  where
a simultaneous {\it re-fit} of the $B$ and
$B^*$ masses from relativistic sum rules
leads to:
$M_b^{PT2}= (4.63 \pm 0.08)$ GeV, while the full-QCD
and HQET sum rules in the large mass limit give the {\it meson-quark mass
difference to two-loops}: $(M_B-M^{NR}_b)_{\infty}
 \simeq (0.6 ^{+0.20}_{-0.10})$ GeV. A
comparison of
these $new$ and $accurate$
results with the existing ones in the literature is
done. As a consequence, the {\it updated} values of
the {\it pseudoscalar decay constants to two-loops} are:
$f_D = (1.37\pm 0.04\pm 0.06)f_\pi$ and $f_B=(1.49\pm 0.06\pm 0.05)f_\pi,
$ which lead to $f_B\sqrt{B_B}=(1.49\pm 0.14)f_\pi$.

\vspace*{0.5cm}
\noindent

\begin{flushleft}
CERN-TH.7405/94 \\
August 1994\\
\end{flushleft}
\vfill\eject
\pagestyle{empty}
\setcounter{page}{1}
\pagestyle{plain}
\section{Introduction} \par
Present accurate measurements of the QCD coupling $\alpha_s$,
mainly from Z-physics and tau-decay data, motivate
a careful reconsideration of the existing estimates of the standard
model parameters. Among others, of a $prime~importance$ is the one of
the quark masses. However, unlike the electron mass, the definition
of the quark masses needs a theoretical framework due to the complication
implied by confinement. In particular, the notion of the pole mass
from perturbation theory, though gauge independent,
appears to be ambiguous, as the summation of large order perturbation
theory induces a non-perturbative term due to renormalon-type
singularities. However, one can still provide a $good~ definition$ of
this pole mass for a truncated QCD series, which can be used
$consistently$ in a given observable known at the same order of
perturbation theory. Unlike the $pole$ mass,
the $running$ mass of the $\overline{MS}$-scheme does not suffer from
this ambiguity, such that its $direct~ estimate$ from the data without
passing through the pole mass should be very useful.

\nin
In this paper, we shall use
 QCD spectral sum rules (QSSR) in order {\it
 to extract, directly} from the data,
and {\it for the first time},
the value of the $running$ heavy quark mass $\overline{m}_Q$\footnote[1]
{However,
an extraction of the running mass using local duality FESR has been
done in \cite{RAF}.}, which is a useful (and well-defined) quantity for
GUTS and some other phenomenology.
 Indeed, many papers have been devoted in the past,
either to the $direct$
estimate of the $Euclidian$ mass $m_Q(p^2=-m^2_Q)$
or/and  of the $pole$ mass $M_Q(p^2=M^2_Q)$
\cite{SVZ}-\cite{SNB}
while the running mass has been
mainly deduced \cite{LEUT,SN1} from its perturbative relation with the pole
mass \cite{BINET}\footnote[2]{For consistency with the two-loop
expression of the two-point correlator, we shall not use the
three-loop
relation in the sum rule analysis.}:
\beq
M_Q(M_Q^2)
=\overline{m}_Q(M_Q^2)\aga 1+\frac{4}{3}\ga\frac{\alpha_s(M_Q^2)}{\pi}\dr
 +K_Q\ga\frac{\alpha_s(M_Q^2)}{\pi}\dr^2 \adr ,
\eeq
where $K_b \simeq 12.4,~K_c \simeq 13.3$ \cite{SCHILCHER}.
Instead, once, we obtain the running mass from the data, we shall use the
previous relation in order to deduce the short-distance
$perturbative$ pole mass $M^{PT}_Q$,
which we shall compare with the
$non$-$relativistic$ pole mass $M^{NR}_Q$ extracted
directly from the data, using non-relativistic sum rules, and with
the mass $M^{nr}_Q$ entering into the non-relativistic
Balmer formula. We shall interpret
the difference between these two values as a measure of the size of the
renormalon effect into the pole mass definition.

\nin
Finally, in comparing our results for the pole masses with the existing ones
in the literature, we shall test the reliability of these previous results
and look for the possible sources of some eventual differences between these
and
the ones in this paper.

\section{Input values of $\alf$ and matching conditions}
We use the value of $\alf(M_Z)$ in the range given in Table 1, where
the new world average is $0.118\pm 0.006$ \cite{BETHKE}, but we have
also considered a slightly higher value of 0.127 in order to be more
conservative. Then, we run  this value until $M_b=4.6-4.7$ GeV, using
the two-loop relation:
\beq
\frac{\alf}{\pi}= a_s \ga
1-a_s\frac{\beta_2}{\beta_1}\log\log(p^2/\al^2) \dr
\eeq
where:
\beq
 a_s= \frac{2}{-\beta_1\log(p^2/\al^2)}
\eeq
and for $n_f$ flavours:
\beq
\beta_1= -\frac{11}{2}+\frac{n_f}{3}~~~~~~\mbox{and}
{}~~~~~~  \beta_2=-\frac{51}{4}+\frac{19}{12}n_f .
\eeq
Following, Ref.\cite{BERN},
we do the matching condition $\alf^{(5)}=\alf^{(4)}$ at this $b$-mass,
in order to extract $\al$ for 4 flavours. We
continue iteratively this procedure for completing Table 1, which is
one of the basic inputs of our analysis. Notice that doing a similar
procedure at the tree-loop level, we reproduce the value of $\al$ given
in \cite{BETHKE}.
\begin{table}[h]
\begin{center}
\begin{tabular}[h]{||lllllll||}
\hline \hline
 & & & & & & \\
$\alf(M_Z)$&$\Lambda_5$[MeV]&$\alf(M_b)$&
$\Lambda_4$[MeV]&$\alf(M_c)$&$\Lambda_3$[MeV]&$\alf(M_\tau)$\\
 & & & & & & \\
\hline \hline
0.112&160 &0.198&233& 0.299&268& 0.269 \\
0.118&225 &0.217&317& 0.350&345& 0.308 \\
0.124&310 &0.240&420& 0.420&433& 0.356 \\
0.127&360 &0.252&475& 0.493&525& 0.407 \\
\hline \hline
\end{tabular}
\caption{\it Value of $\alf$ and $\Lambda$ to two-loops at different scales and
\end{center}
\end{table}
 \section{ Running masses from the $\Psi$- and
 $\Upsilon$-systems}

In so doing, let us consider
the two-point vector correlator:
\bea
\Pi^{\mu\nu}_b(q^2,M^2_Q) &\equiv& i \int d^4x ~e^{iqx} \
\la 0\vert {\cal T}
J^{\mu}_Q(x)
\ga J^{\nu}_Q(o)\dr ^\dagger \vert 0 \ra \\ \nnb
&=& -\ga g^{\mu\nu}q^2-q^\mu q^\nu \dr \Pi_Q(q^2,M^2_Q),
\eea
where $J^{\mu}_Q(x) \equiv \bar Q \gamma^\mu Q (x)$ is the local vector
current of the heavy quark $Q$. The correlator obeys the well-known
K\"all\`en-Lehmann dispersion relation:
\beq
\Pi_Q(q^2,M^2_Q) = \int_{4M^2_Q}^{\infty} \frac{dt}{t-q^2-i\epsilon}
{}~\frac{1}{\pi}~\mbox{Im}  \Pi_Q(t) \ \ \ + \ \ \ \mbox{subtractions},
\eeq
which expresses in a clear way the {\it duality} between the spectral
function Im$ \ \Pi_Q(t)$, which can be measured experimentally, as here
it is related to the $e^+e^-$ into $\Psi$ or $\Upsilon$-like states total
cross-section or their leptonic widths as:
\beq
\mbox{Im}  \Pi_Q(t) =\frac{3}{4\alpha^2}\frac{1}{Q^2_Q}
\sum_{i}
{\Gamma_i M_i} \delta (t-M^2_i)~~+~~
\Theta (t-t_c) \mbox{Im} \Pi^{QCD}_Q(t).
\eeq
$Q_Q$ is the heavy quark charge in units of e;
$\Gamma_i$ is the electronic width of the resonances with the value
given in PDG 92 \cite{PDG}; $t_c$ is the QCD continuum threshold
which we fix just above the last known radial excitation which is
respectively about (5 GeV$)^2$ and (12 GeV$)^2$ for the $\Psi$ and
$\Upsilon$ families (however, it should be noticed that your result
is not dependent on this choice due to the
almost complete dominance of the lowest ground state at the stability
point).
$\Pi_Q(q^2,M^2_Q)$ can be calculated directly in
QCD, even at $q^2=0$,
provided that $M^2_Q-q^2 \gg \Lambda^2$. For the perturbative part,
we shall use (without expanding in 1/M)
the Schwinger extrapolation formula to two-loops:
\beq
\mbox{Im} \Pi^{pert}_Q(t)
\simeq \frac{3}{12\pi}v_Q\ga \frac{3-v^2_Q}{2} \dr
\aga 1+\frac{4}{3}\alf f(v_Q) \adr ,
\eeq
where:
\beq
v_Q= \sqrt{1-4M^2_Q/t}
\eeq
and:
\beq
f(v_Q)=\frac{\pi}{2v_Q}-\frac{(3+v_Q)}{4}\ga \frac{\pi}{2}-\frac{3}{4\pi}
\dr
\eeq
is the Schwinger function \cite{SCHWI}. We express this
spectral function in terms of the running mass by using Eq. (1)
to two-loops and the
$\alf\log(t/M^2_Q)$-term appearing in Eq. (1) for off-shell quark.
We shall add to this perturbative
expression the lowest dimension $\la \alf G^2 \ra $
non-perturbative effect (it is known \cite{HEAVY} that, for a heavy-heavy quark
correlator, the quark condensate contribution
is already absorbed into the gluon
one)
which among the available higher dimension
condensate-terms can only give a non-negligible contribution. We shall
use the range of values:
\beq
\la \alf G^2 \ra = (0.06 \pm 0.03)~ \mbox{GeV}^4,
\eeq
from different QSSR analysis \cite{SNB}; a value confirmed by the recent
ALEPH measurement of this quantity from tau-decay data \cite{ALEPH},
which, at the same time,
exclude higher values of this condensate advocated sometimes
in the literature.

\nin
QSSR is an improvement of the previous
dispersion relation. For our purpose, we shall consider the ratios:
\beq
{\cal R}_n \equiv \frac{{\cal M}^{(n)}}{{\cal M}^{(n+1)}}~~~~~~~
\mbox{and}~~~~~~~
{\cal R}_\tau \equiv -\frac{d}{d\tau} \log {{\cal L}},
\eeq
and their finite energy sum rule (FESR) variants, which come from
the moment sum rules
(finite number of derivatives and finite values of $q^2$):
\beq
{\cal M}^{(n)} \equiv \frac{1}{n!}\frac{\partial^n \Pi_Q(q^2)}
{\ga \partial q^2\dr^n} \Bigg{\vert} _{q^2=0}
= \int_{4M^2_Q}^{\infty} \frac{dt}{t^{n+1}}
{}~\frac{1}{\pi}~ \mbox{Im}  \Pi_Q(t),
\eeq
or infinite number of derivatives and infinite values of $q^2$, but
keeping their ratio fixed as $\tau \equiv n/q^2$
(Laplace or Borel or exponential sum rules):
\beq
{\cal L}(\tau,M^2_b)
= \int_{4M^2_Q}^{\infty} {dt}~\mbox{exp}(-t\tau)
{}~\frac{1}{\pi}~\mbox{Im} \Pi_Q(t).
\eeq
The ratios of sum rules are
more appropriate for the estimate of the
quark mass as these ratios equate $directly$ the mass squared of
ground state to that of the quark. They also eliminate, to leading order,
some artefact dependence due to the sum rules (exponential weight factor
or number of derivatives).

\nin
In principle, the pairs $(n,t_c)$, $(\tau,t_c)$ are free external
parameters in the analysis, so that the optimal result should be
insensitive to their variations. Stability criteria, which are equivalent
to the variational method, state that the optimal results should
be obtained at the minimas or at the inflexion points in $n$ or $\tau$,
while stability in $t_c$ is useful to control the sensitivity of the
result in the changes of $t_c$-values (in the present case of the
$\Psi$ and $\Upsilon$ systems, $t_c$-stability is manifest due to the
negligible effect of the QCD continuum).
Stability criteria have also been tested in
models such as
harmonic oscillator, where the exact and approximate
solutions are known \cite{BERT}. The {\it most conservative
optimization criteria} which include various types of optimizations
in the literature are the following: the
optimal result is obtained in the region,
from the beginning of $\tau / n$ stability (this corresponds in most
of the cases to the so-called plateau discussed often in the literature,
but in my opinion, the interpretation of this nice plateau as a good
sign of a good continuum model is not sufficient, in the sense
that the flatness of the
curve extends in the uninteresting high-energy region where the
properties of the ground state are lost),
until the beginning of the $t_c$
stability, where the value of $t_c$ corresponds to about the one fixed by
FESR duality constraints.
The earlier {\it sum rule window} introduced by SVZ, stating that the
optimal result should be in the region where both the non-perturbative
and continuum contributions are {\it small} is included in the previous
region.
 Indeed, at the stability
point, we have an equilibrium between the continuum and non-perturbative
contributions, which are both small,
while the OPE is still convergent  at this point.

\nin
The gluon condensate contribution to the moments ${\cal M}^{(n)}$ and
so to ${\cal R}_n$ can be copied from the original work of SVZ \cite{SVZ}
and reads:
\beq
{\cal M}^{(n)}_G=-{\cal M}^{(n)}_{pert} ~\frac{(n+3)!}{(n-1)!(2n+5)}
\frac {4\pi}{9}\frac{\la \alf G^2 \ra}{\ga 4M_Q^2 \dr^2},
\eeq
where ${\cal M}^{(n)}_{pert}$ is the lowest perturbative expression
of the moments.
The one to the Laplace ratio ${\cal R}_\tau $ can be also
copied from the
original work of Bertlmann \cite{BERT}, which has been expanded
recently in $1/M_Q$ by \cite{DOM}. It reads:
\beq
{\cal R}^G_\tau \simeq (4M^2_Q)\frac{2\pi}{3}\la \alf G^2 \ra \tau^2
\ga 1+\frac{4}{3\omega}-\frac{5}{12\omega^2} \dr ,
\eeq
where $\omega = 4M^2_Q \tau$. Due to the poor accuracy of $\la \alf G^2 \ra$,
we
used the recent two-loop calculation of \cite{BROAD2}.

\nin
We give the results of our analysis in
Fig. 1 from the FESR version of these relativistic
sum rules as we have transferred,
into the QCD side, the QCD continuum contribution, such that in the
ratio, we only remain with the resonance contributions in the RHS
of the sum rule. At the stability point or for large n-values, the
experimental ratio of the sum rules give, with an accuracy better
than 1$^0/_{00}$, the mass squared of the lowest ground state. The
curves in Fig. 1 are
for a given value of $\alf (M_b)=0.217$ and the corresponding
$\alf (M_c)=0.350$. As explained previously, our best solution
is obtained at the minimum of the curve for ${\cal R}_\tau$, while
for ${\cal R}_n$, the curve is flat and insensitive to n. As one can
notice, the reproduction of the resonance mass needs a sharp value of
the running mass, while the moments and Laplace sum rules give
practically the same solution of $\overline {m}_b$. We have also
checked that the result is almost insensitive to the choice of the
subtraction point varying in the range from $\tau^{-1/2}\simeq 1.4$ GeV
to $M_Q$. That is due to the small effect of the radiative corrections
in the sum rule analysis. However, the natural choice of the scale
as dictated by the RGE is $M_b$ for the moments \cite{SNY}
and $\tau^{-1/2}$ for
the Laplace \cite{RAF2}.
It is also clear
due to the smallness of the $c$-quark mass that we cannot use at our
approximation the moments for this channel.
 \section{Non-relativistic
 pole masses from the $\Psi$- and $\Upsilon$-systems}
Let us now extract directly the {\it non-relativistic}
pole masses from the $\Psi$ and $\Upsilon$
data, by using the non-relativistic version of the
previous sum rules. This determination is interesting as a comparison of this
result with the {\it short distance perturbative pole mass}
 from the perturbative expression in Eq. (1) via the running
mass can measure the size of the renormalon contribution discussed in
\cite{REN}

\nin
The non-relativistic Laplace sum rule has been discussed in details
by Bell and Bertlmann \cite{BERT}. In this case the non-relativistic
ratio of moments ${\cal R}^{NR}_{\tau}$ is about $\sqrt{{\cal R}_\tau}$ and
can be deduced from ${\cal R}_\tau$ by the introduction of the non-relativistic
variable:
\beq
\tau_{NR}= 4M_Q\tau .
\eeq
The results of this analysis are shown in Fig. 2 for a given value of $\alf$
and
different values of $M_Q$, where as in Fig. 1, the agreement with the data
gives
constraint on the value of $M_Q$.

\nin
The non-relativistic version of the moments has been
discussed in \cite{VOL,VOL2}, where a summation of the higher order Coulombic
co
has been done. The moment reads \cite{VOL}:
\bea
{\cal M}^{(n)}_{NR} &\equiv &
\int_{4M^2_b}^{\infty} ~dt~ \mbox{exp}\ga \ga1-\frac{t}{4M^2_b}\dr n\dr
\frac{1}{\pi}~\mbox{Im}~\Pi_b(t) \nnb \\
 & = & M^2_b~\ga \frac{3}{4\pi^2} \dr
\frac{\sqrt{\pi}}{n^{3/2}}\ga 1-\frac{16}{3}\frac{\alf}{\pi}\dr\aga
\Phi_s(\gamma)-\frac{\pi}{72}\la \alf G^2\ra \frac{n^3}{M^4_b}X_s(\gamma)\adr
,
\eea
where:
\beq
\gamma = \frac{2}{3} \alf n^{1/2} ~~~~~~~~X_s(\gamma)\simeq
e^{-0.8\gamma}\Phi_s
\eeq
and:
\beq
\Phi_s(\gamma)=1+2\sqrt{\pi}\gamma+\frac{2\pi^2}{3}\gamma^2+4\sqrt{\pi}\sum_{i=1
\ga \frac{\gamma}{n} \dr ^3 \mbox{exp}\ga\frac{\gamma}{n}^2\dr\ga 1+\mbox{erf}
\ga\frac{\gamma}{n}\dr\dr
\eeq
with:
\beq
\mbox{erf}(x)= \frac{2}{\sqrt{\pi}}\int_{0}^{x}dt~\mbox{exp}\ga -t^2\dr .
\eeq
The result from this sum rule is also shown in Fig. 2, where we anticipate that
the value $\approx$ 4.8 GeV claimed in \cite{VOL} is not reproduced,
even if we use the same input parameters as \cite{VOL}, though we agree with
the
value of $n \approx 25$ and on the corresponding value of
$\Phi_s(\gamma)\approx
Instead, our result is more
similar to the previous value obtained by the one of the authors \cite{VOL2},
from a similar sum rule.
\section{Conclusions from the $\Psi$- and $\Upsilon$-systems}
We give our complete results  in Table 2 for the $b$ and in
Table 3 for the $c$.

\begin{table}[h]
\begin{center}
\begin{tabular}[h]{||l|llllll||}
\hline \hline
 & & & & & &\\
$\alf(M_b)$&$\overline
{m}_b (M_b)$&$\hat {m}_b$&
$m^{EU}_b$&$M_b^{PT2}$&$M_b^{PT3}$&$M_b^{NR}$\\
 & & & & & &\\
\hline \hline
0.198 &4.26 &8.33& 4.25& 4.62 &4.83&4.68 \\
0.217 &4.23 &7.83& 4.21& 4.62 &4.87&4.69 \\
0.240 &4.19 &7.23& 4.17& 4.62 &4.92&4.71 \\
0.254 &4.17 &7.05& 4.15& 4.62 &4.96&4.71  \\
\hline \hline
\end{tabular}
\caption{\it Two-loop values of the $b$-quark masses for different values of
$\a
for 5 flavours except for $M^{PT}_b$ given also at the tree-loop level.}
\end{center}
\end{table}

\nin

\begin{table}[h]
\begin{center}
\begin{tabular}[h]{||l|llllll||}
\hline \hline
 & & & & & &\\
$\alf(M_c)$&$\overline {m}_c (M_c)$&$\hat {m}_c$&
$m^{EU}_c$&$M_c^{PT2}$&$M_c^{PT3}$&$M_c^{NR}$\\
 & & & & & &\\
\hline \hline
0.299 &1.25 &1.78& 1.24& 1.41&1.56 &1.42 \\
0.350 &1.23 &1.60& 1.22& 1.41&1.62 &1.44 \\
0.420 &1.19 &1.39& 1.18& 1.40&1.69 &1.46 \\
0.493 &1.16 &1.23& 1.15& 1.40&1.78 &1.49  \\
\hline \hline
\end{tabular}
\caption{\it Two-loop values of the $c$-quark masses for different values of
$\a
4 flavours except for $M^{PT}_c$ given also at the tree-loop level.}
\end{center}
\end{table}

\nin
The running mass $\overline{m}_Q$ directly obtained from
the sum rules for different
values of $\alf$ (first column) from Table 1,
is given in the second column. We use the two-loop relation \cite{FLO,SNB}:
\beq
\overline{m}_Q(p^2) = \hat{m}_Q\ga -\beta_1\frac{\alf(p^2)}{\pi}\dr
^{-\gamma_1/\beta_1} \aga 1~+~\frac{\beta_2}{\beta_1}
\ga \frac{\gamma_1}{\beta_1}-\frac{\gamma_2}{\beta_2}\dr
\ga \frac{\alf}{\pi}\dr
\adr ,
\eeq
where $\hat{m}_Q$ is the invariant mass \cite{FLO}(third column); $\gamma_1=2$
and $\gamma_2= 101/12-5n_f/18$
(third column). The gauge-dependent Euclidian mass
$m^{EU}_Q $ (fourth column) reads in the Landau gauge:
\beq
m^{EU}_Q(p^2=-M^2_Q)
=\overline{m}_Q(M_Q)\aga 1+\ga\frac{\alpha_s(M_Q)}{\pi}\dr
\ga -2\log2 +\frac{4}{3} \dr \adr .
\eeq
We use Eq. (1) to get the short distance {\it perturbative} pole mass
$M^{PT}_Q$
and 6th columns).
The pole mass $M^{NR}_b$ directly obtained from non-relativistic sum rules is
given in the last column. We have only considered the one from the Laplace
ratio
${\cal R}^{NR}_{\tau}$, which appears to be more reliable than the
moment  analysis.

\nin
 Using the {\it world average}\cite{BETHKE}: $\alf (M_Z)=0.118 \pm 0.006$,
we obtain, for the $b$-quark mass at the two-loop level, the values, in units
of
\beq
\overline{m}_b(M_b)= 4.23 ~^{+0.03}_{-0.04}~\pm 0.02 ~~~
\hat{m}_b=7.83~^{+0.40}_{-0.60}~\pm 0.04 ~~~
m^{EU}_b=4.21\pm 0.04 \pm 0.02
\eeq
and:
\beq
M^{PT2}_b=4.62 \pm 0.02 ~~~~~~~~~~~~~~~
M^{NR}_b=4.69~^{-0.01}_{+0.02} \pm 0.02.
\eeq

\nin
It is informative to compare this two-loop result with the ones in the existing
which are known at the same level of accuracy.
The value of the invariant mass $\hat {m}_b$ agrees with the one given in
\cite{
while the error is reduced here by a factor of about 2 due to the progress in
th
measurement of $\alf$. The value of the Euclidian mass $m^{EU}_b$ obtained here
the running mass, is compatible with the one extracted directly from the sum
rul
\cite{SVZ}-\cite{BERT}, despite the different values of $\alf$ used in these
pap
which most of  the time are incompatible with the range given
in Table 1. Indeed, the use
of the Euclidian mass in the sum rules, minimizes the size of the radiative
corr
and then the one of $\alf$ in the sum rule estimate of $m^{EU}$. However,
the uncertainty due
to $\alf$ appears when one translates this mass into the pole one $M^{PT2}_b$,
t
\beq
M^{PT2}_Q=m^{EU}_Q(p^2=-M^2_Q)
\aga 1+2\log2\ga\frac{\alpha_s(M_Q)}{\pi}\dr
\adr ,
\eeq
or if one extracts directly the pole mass from relativistic sum rules. One can
a
notice that $M^{PT2}_Q$ is almost constant in the given
range of $\alf$ values, as the change in the running mass is compensated by the
in $\alf$ via Eq. (1). The value of $M^{NR}_b$ agrees
with the one of \cite{BERT,DOM}, while
we have improved the recent re-estimate of \cite{DOM}
by using consistently the value
of $\Lambda$ and the corresponding two-loop expression
 of $\alf$. However, as already stressed, our
analysis using the same non-relativistic
moments in Eq. (18) does not favour the
value $4.8$ GeV claimed by \cite{VOL}.

\nin
It is interesting to compare $M^{NR}_b$, with the {\it dressed mass}
 $M^{nr}_b$ from the
non-relativistic treatment  of the bound state motion, where the leading
non-per
effect is included. For a sufficiently heavy quark, the {\it Balmer formula}
for the lowest ground state reads \cite{GASSER,LEUT}:
\beq
M_{\Upsilon}=2M^{nr}_b\aga 1-\frac{1}{8}
\ga C_F\alf \dr ^2+ {\cal O} \alf^3+
\frac{\pi}{2}~\frac{\la{\alf G^2}\ra}{\ga M^{nr}_b \dr ^4}
\frac{\ga \epsilon_{10}\simeq 1.468 \dr}{\ga C_F\alf \dr ^4} ~ \adr
\eeq
where $C_F =4/3$, from which one
deduces\footnote[3]{The next-to-leading corrections
to this formula have been evaluated in \cite{YND},
 but we have not included them in our estimate in Eq. (28)
in order to have a consistent comparison
with $M^{NR}_b$ obtained at the two-loop level. These higher order
corrections are taken into account in Eq. (29).}:
\beq
M^{nr}_b = (4.57\pm 0.09\pm 0.14)~\mbox{GeV},
\eeq
 where the errors are due to $\alf$ and $\la \alf G^2 \ra $.
 The difference of the central value
with the one ($4.76 \pm 0.10 \pm 0.05$) GeV
obtained in \cite{LEUT,YND} is mainly due to the different value of
$\alf$ and of the gluon condensate used there. However, the error
 due to the latter has been underestimated in \cite{YND}. The large
role of the gluon condensate also indicates that the $b$-quark is not
enough heavy, such that we have not yet reached the coulombic regime.

\nin
One might
also follow \cite{YND} by including higher order $\alf$
corrections and by trying to optimize the
convergence of the QCD series, which can be
realized at a scale $\mu \approx 1.5 $ GeV. In this way,
the gluon condensate effect is almost negligible, while the perturbative series
dominated by the $\alf^2$ and $\alf^4$ terms. However, the price to pay is the
s
of the result on the $\alf$-value. One obtains in this way:
\beq
M^{nr}_b = (4.94\pm 0.10 \pm 0.03)~\mbox{GeV},
\eeq
where, we again notice the underestimate of the error in \cite{YND}.
One might expect from its derivation based on the $\overline{b}b$
potential, that $M^{nr}_b$ is likely similar to the
so-called {\it constituent mass} used in the potential model
approach for predicting the mass spectra \cite{RICH}, while it
is quite unlikely to deduce the running mass
of the $\overline{MS}$-scheme from it. It is interesting to compare this value
with the short-distance pole mass deduced from
the three-loop relation in Eq. (1),
which reads:
\beq
 M^{PT3}_b=(4.87 \pm 0.05 \pm 0.02) ~ \mbox {GeV}.
\eeq
It is important to notice that the inclusion of the three-loop terms into the
re
in Eq. (1) has boosted appreciably the value of
the pole mass $M^{PT3}_Q$, such that the next improvment of this estimate
is the inclusion of the unknown $\alf^3$-term \footnote[4]{The estimate
of this effect \`a la \cite{KATA}
cannot be used here in a straightforward way, as the running
mass has an anomalous dimension. However, if we use a guessed estimate
of the coefficient based on a geometric sum \`a la \cite{BRAT}, we
would obtain an effect of $\pm {\cal O} 100 (\alf/\pi)^3 \approx 3\%$
for the $b$, which might be an overestimate.} as well as the $\alf^2$-
term in the hadronic correlator.

\nin
 One has also to notice that one should be careful in the use
of this three-loop
mass in different phenomenological applications. In particular,
for the case of different sum rule analysis
(estimate of the decay constants,...),
where the hadronic correlators are only known to two-loop accuracy,
one {\it should neither
use} $M^{PT3}_Q$, $nor$ $M^{nr}_b$ for consistency.

\nin
A comparison of $M^{PT}_b$ with $M^{NR}_b$ at the two-loop level leads to:
\beq
\Delta M_b\equiv M^{NR}_b-M^{PT2}_b = (70 ~^{-10}_{+20} )~ \mbox{MeV},
\eeq
while a comparison of $M^{nr}_b$ (which we expect from its
definition to be identical with $M^{NR}_b$)
with the value of the short-distance perturbative mass at the three-loop
level:
\beq
\Delta M_b\equiv M^{nr}_b-M^{PT3}_b =(70 \pm 100)~ \mbox{MeV},
\eeq
gives about the same small value but with a larger error.
 We might expect that
this mass difference measures
the size of the renormalon singularities, which
after a resummation of the QCD large order perturbative series is expected to
give an effect of the order of $\Lambda$
\cite{REN}\footnote[5]{ However, a precise quantitative derivation of
that value from the theoretical point of view is still lacking as one has to
wor
limit of
large number of flavours, where one loses asymptotic freedom.}.
Our analysis
indicates that the renormalon effect can be even smaller than the
present value  of $\Lambda$ given in Table 1.

\nin
For the charm quark, our results, in units of GeV, are:
\beq
\overline{m}_c(M_c	)= 1.23 ~^{+0.02}_{-0.04}~\pm 0.03 ~~~
\hat{m}_c=1.60~^{+0.18}_{-0.21}~\pm 0.04 ~~~
m^{EU}_c=1.22~^{+0.02}_{-0.04}\pm 0.03
\eeq
and:
\beq
M^{PT2}_c=1.41 \pm 0.03 ~~~~~~~~~~~~~
M^{NR}_c=1.44\pm 0.02 \pm 0.03,
\eeq
where the errors are respectively due to $\alf$
and to the gluon condensate.
At the three-loop level, we have:
\beq
M^{PT3}_c=(1.62\pm 0.07  \pm 0.04)~\mbox{GeV},
\eeq
where the last error is due to an estimate of the $\alf^3$-unknown
correction.
The same
discussions as for the $b$-quark case are valid here, when we
do the comparison with
previous estimates given in the literature.
A comparison between $M_c^{PT}$ and $M_c^{NR}$
shows that:
\beq
\Delta M_c \simeq (30 \pm 20 ) ~\mbox{MeV},
\eeq
which may indicate that the renormalon effect is smaller
for the charm than for the  bottom
quark. It may suggest that the leading-order (in $1/n_f$ and
most probably in $1/M_Q$) calculation of this contribution might be
 largely affected by the non-leading corrections.
\section{ Quark masses from the heavy-light mesons}
We re-examine in the same way, the extraction of the $b$-quark mass from
the $B$ and $B^*$ meson masses, in order to test the stability of the
result\cite{SNX}:
\beq
M_b^{PT2}= (4.56 \pm 0.05)~\mbox{GeV},
\eeq
from the ratio of relativistic
sum rules. For convenience, we introduce the variable $E_c$:
\beq
t_c= (E_c+M_b)^2.
\eeq
In order to be conservative with the different choices in the literature,
we use:
\beq
E_c = (1.3 \pm 0.3)~\mbox{GeV}.
\eeq
by multiplying by a factor 3 the error given in \cite{SNZ}. We give the
domain of predictions versus $M_b^{PT2}$ in Fig. 3,
from which we conclude:
\beq
M_b^{PT2}=(4.63 \pm 0.08)~\mbox{GeV},
\eeq
which is consistent with the previous value and with the one
from the $\Upsilon$-systems. The large error is due to the conservative
range of $E_c$, while the smaller central value in Eq. (37) is related
to the higher value of $E_c$ and to the small
value of $\alf$ used in previous
works.

\nin
We have also performed the analysis using nonrelativistic sum rules.
Though the result for $M_b^{NR}$ is consistent with the one from
the vector channel, it is quite influenced by the value of $\alf$ or
$\Lambda$ used, which renders the analysis quite inaccurate. That is
partly due to a larger value of the $\alf$ correction in this sum rule
and to the disappearance of the stability points in the analysis.
Finally, we have compared the previous results on $M_b^{NR}$
with the one from
heavy quark effective theory (HQET) sum rule \cite{BALL} in the infinite
mass limit, which gives:
\beq
\delta M^{\infty}
_b\equiv (M_B-M^{NR}_b)_{\infty} \simeq (0.5-0.6)~\mbox{GeV},
\eeq
while the
usual full-QCD relativistic sum rule for
increasing value of $M^{NR}_b~(\geq 5$ GeV) gives \cite{SNY}:
\beq
\delta M^{\infty}_b \simeq (0.6-0.8)~\mbox{GeV},
\eeq
from which, we deduce the conservative value:
\beq
\delta M^{\infty}_b \simeq (0.6^{+0.2}_{-0.1})~\mbox{GeV}.
\eeq
These results are in agreement with the previous determinations from
the $\Upsilon$-systems,
though they are less accurate. However, the agreement of the value
of $M^{NR}_b$ from the heavy-heavy and heavy-light systems may
indicate that the ambiguity due to the renormalon effects into the
definition of the pole masses entering into the two channels is
negligible, which in some sense is supported by the small value of
the mass-difference $M_b^{NR}-M_b^{PT2}$.
\section{Conclusions}
We have updated the theoretical estimate of the heavy quark masses
from QCD spectral sum rules,
by using the latest average value of $\alf$.
The best results come from the $\Upsilon$ and $\Psi$ data
and are given to two-loops in Eqs. (24), (25), (33) and (34), while the
three-loop results for the short distance pole masses
are in Eqs. (30) and (35).
The good accuracy from the vector channels is
somewhat expected as the data are known in details there.
Results from the $B$ and $B^*$ mesons are less accurate, which are
mainly due to the lesser control of the QCD continuum threshold,
which has been taken in a large range of values, in order to have
a more conservative result.

\nin
We have resolved the apparent disagreement of the results given in the
literature by a careful re-examination of each analysis. One of the
main sources of discrepancies is due to
the different values of $\alf$ used
in the literature. Our value of the Euclidian mass is in good agreement
with the direct determinations in \cite{SVZ}-\cite{BERT}. Our value
for the running mass agrees within the errors with the previous results
\cite{LEUT,SN1} deduced through the perturbative pole masses.
The small discrepancy in this determination
can be resolved if one uses the
same value of $\alf$. Our value of the short distance pole mass agrees
from the direct determination \cite{BERT} for a given value of $\alf$.
The apparent discrepancy with \cite{REINDERS} disappears once one uses
the same value of $\alf$.
Our two-loop result for the non-relativistic mass agrees with the
direct determination \cite{BERT,DOM} when one uses in these papers the
same value of $\alf$ to two-loop accuracy. However,
we do not $exceptionnally$ recover the
value 4.8 GeV of \cite{VOL} but his earlier result \cite{VOL2}.
Moreover, our two-loop result rules out the value of the pole charm quark
mass of 1.35 GeV and of the $b$-quark pole mass of 4.8 GeV,
used {\it sometimes abusively} in the literature.

\nin
The use of the three-loop relation in Eq. (1)
suggests that radiative corrections are relatively large and shifts the
two-loop pole mass by +3.8 and +14 $\%$ for the $b$ and $c$ quark masses.
This effect is, however, interesting as it restores the agreement between
the pole mass with the value of the so-called constituent mass
favoured in potential models \cite{RICH}, which is likely to be the
same as $M^{nr}_b$ appearing in the Balmer formula, with the value
given in Eq. (29).
In particular, the {\it b-c} mass difference which appears to be a safe
prediction of potential models is obtained within the errors
from our analysis. The three-loop value of the charm pole mass
is also in agreement with
the recent lattice calculations including $\alf^2$-corrections
\cite{LATTICE}, while the value of $M^{nr}_b$ including higher
order corrections, should be similar to the one obtained recently
from the lattice results in \cite{LATT}, due to the non-relativistic
way used in the lattice calculation. However, a
sharp comparison still needs
lattice results with smaller error bars and
in the $unquenched~approximation$, though one expects the probability
for creating  $\bar qq$ pairs to be small.
\nin
Finally, our results suggest that the mass difference between the
short-distance and the non-relativistic pole masses is small of
about 70 MeV, and presumably less. This is an indication that the
non-perturbative effects induced by the summation of the QCD series
at large order due to renormalon-type singularities can be relatively
negligible.

\nin
One of the immediate consequences of these $new$ two-loop results is the
prediction on the $D$ and $B$
decay constants which are very sensitive to the value
of the quark masses used in the analysis, whose explicit mass-dependence
has been studied in \cite{SNW}. With the previous values of the short
distance pole mass which are consistent with the uses of the relativistic
Laplace sum rule results, one can deduce from the table of \cite{SNW},
the updated and improved values:
\beq
f_D = (1.37\pm 0.04\pm 0.06)f_\pi~~~~~~~~~~~~~
f_B=(1.49\pm 0.06\pm 0.05)f_\pi,
\eeq
where the errors come respectively from the
pseudoscalar sum rule analysis
and from the quark masses.
 We consider this result as an improvement of
the one in \cite{SNY}, where due to the change in the quark-mass values,
the value of $f_D$ (resp. $f_B$) has slightly increased
(resp. decreased). By combining this result with the bag factor
$B_B=1\pm 0.15$ obtained to two-loops in \cite{PIVO}, one can deduce:
\beq
f_B\sqrt{B_B}=(1.49\pm 0.14)f_\pi,
\eeq
after adding the different errors quadratically. The use of this value
should improve the phenomenological constraints on the values of the
CKM-mixing parameters (see e.g. \cite{ALI}).

\vfill \eject
\section*{Figure captions}
{\bf Fig. 1:} {\bf a)}:
value of $M_\Upsilon$ from the ratio of the relativistic Laplace sum
rule ${\cal R}_\tau^b$
versus the sum rule variable $\tau$
for different values of the running mass $\overline{m}_b(M_b)$ given
$\alf(M_b)=0.217$; {\bf b)}: analogous to {\bf a)} but from the ratio of
moments; {\bf c)}: analogous to {\bf a)} but for the $\Psi$.
\vspace*{0.5cm}

\nin
{\bf Fig. 2:} {\bf a)}: non-relativistic analogue of Fig. 1a), with
$\tau_{NR}=4M_b\tau$. The dashed line is the experimental data which
tend to $M_\Upsilon$ for large $\tau_{NR}$; {\bf b)}: confrontation
of the two sides of the non-relativistic moment in Eq. (18) versus $n$
and for different values of the pole mass, given $\alf(M_b)$=0.217.
The dashed lines are the LHS from the data; {\bf c)}: analogous to
{\bf a)} but for the $\Psi$.
\vspace*{0.5cm}

\nin
{\bf Fig. 3:} Simultaneous fit of the $B$ and
$B^*$ masses from the ratios
of the relativistic moments. The horizontal
 lines correspond to the masses
from the data. The corresponding theoretical domains are for $E_c = (1.3
\pm 0.3)$ GeV for different values of the pole mass from the
$n$-stability.
\vfill \eject

\end{document}